\documentclass{article}

\usepackage{arxiv}

\usepackage[utf8]{inputenc} 
\usepackage[T1]{fontenc}    
\usepackage{hyperref}       
\usepackage{url}            
\usepackage{booktabs}       
\usepackage{amsfonts}       
\usepackage{nicefrac}       
\usepackage{microtype}      
\usepackage{lipsum}

\usepackage{amsmath,amsfonts,amssymb}
\usepackage{algorithm}
\usepackage[noend]{algpseudocode}
\usepackage{verbatim}
\usepackage{gensymb}
\usepackage{textcomp}
\usepackage{pdfpages}
\usepackage{graphicx}
\usepackage[space]{grffile}
\newcommand{\wave}{\raise.17ex\hbox{$\scriptstyle\sim$}}
\usepackage{tikz}

\usepackage{flushend} 
\usepackage[caption=false]{subfig}
\usepackage{enumitem}
\usepackage{rotating}
\usepackage{natbib}
\usepackage{multicol}

\title{Fast and Accurate Reconstruction of Pan-Tilt RGB-D Scans via Axis Bound Registration}

\author{
Jung-Hyun Byun\\
  Department of Computer Science\\
  Yonsei University, Korea\\
  \texttt{junghyun.ian.byun@gmail.com} \\
   \And
Tack-Don Han \\
  Department of Computer Science\\
  Yonsei University, Korea\\
  \texttt{hantack55@gmail.com} \\
}

\begin{document}
\maketitle

\begin{abstract}
A fast and accurate algorithm is presented for registering scans from an RGB-D camera on a pan-tilt platform.
The pan-tilt RGB-D camera rotates and scans the entire scene in an automated fashion. 
The proposed algorithm exploits the movement of the camera that is bound by the two rotation axes of the servo motors so as to realize fast and accurate registration of acquired point clouds. 
The rotation parameters, including the rotation axes, pan-tilt transformations and the servo control mechanism, are calibrated beforehand. 
Subsequently, fast global registration can be performed during online operation with transformation matrices formed by the calibrated rotation axes and angles. 
In local registration, features are extracted and matched between two scenes. 
For robust registration, false-positive correspondences are rejected based on their distances to the rotation trajectories.
Then, a more accurate registration can be achieved by minimizing the residual distances between correspondence pairs, while estimated transformations are bound to the rotation axes. 
Results of comparative experiments validate that the proposed method outperforms state-of-the-art algorithms of various approaches based on camera calibration, global registration, and simultaneous-localization-and-mapping in terms of root-mean-square error and computation time.
\end{abstract}

\section{Introduction}

\begin{multicols}{2}
Indoor scene reconstruction is of crucial importance in various research areas, in which 3D geometry data is required, such as robotics, architecture, and augmented reality.
More specifically, in robotics, 3D map data is required for a robot to localize itself, plan a path, and navigate to a specific location as ordered. The practice of this localization and mapping is often abbreviated as SLAM (Simultaneous Localization and Mapping). 
In architecture, indoor scenes are reconstructed, so that users can navigate virtually, and are then exported as CAD files for users to correct reconstruction errors and further edit to meet their needs  \citep{ikehata2015structured}. 
In conjunction with augmented reality (AR), reconstructed models are used as surface geometry for projection-based AR, where geometric distortions are corrected to deliver pure immersive AR experience \citep{wilson2012steerable}.

In general, RGB depth (RGB-D) cameras are hand-held by the user and are moved freely around the environment to map and reconstruct the geometry of indoor scenes, particularly for SLAM applications \citep{sturm12iros}.
However, in various cases, RGB-D cameras are attached to pan-tilt rotating platforms and collect 3D data captured in pan-tilt sweeps \citep{wilson2012steerable, naweed2014enhancement, ambrucs2014meta}.
Using pan-tilt platforms with RGB-D cameras has several advantages.
Firstly, the field-of-view of the cameras is limited, when a wide range of data should be captured from the environment.
In such a case, steering the camera with a pan-tilt unit would be preferred  \citep{niu2017calibration}, over noisy hand-held alternatives.
Secondly, the detection range and resolution of the RGB-D camera may be limited to capture the geometry of even a standard-sized room.
Thus, given that the scene is stationary, using a pan-tilt RGB-D camera to capture the room from its center, would be an ideal option \citep{naweed2014enhancement}.
Lastly, as the scanning and capturing process can be automated and computer-controlled \citep{ambrucs2014meta}, it would be more convenient and accurate to use pan-tilt RGB-D cameras.
\end{multicols}

\twocolumn

\begin{figure}[!htb] 
\centering
\includegraphics[width=\linewidth]{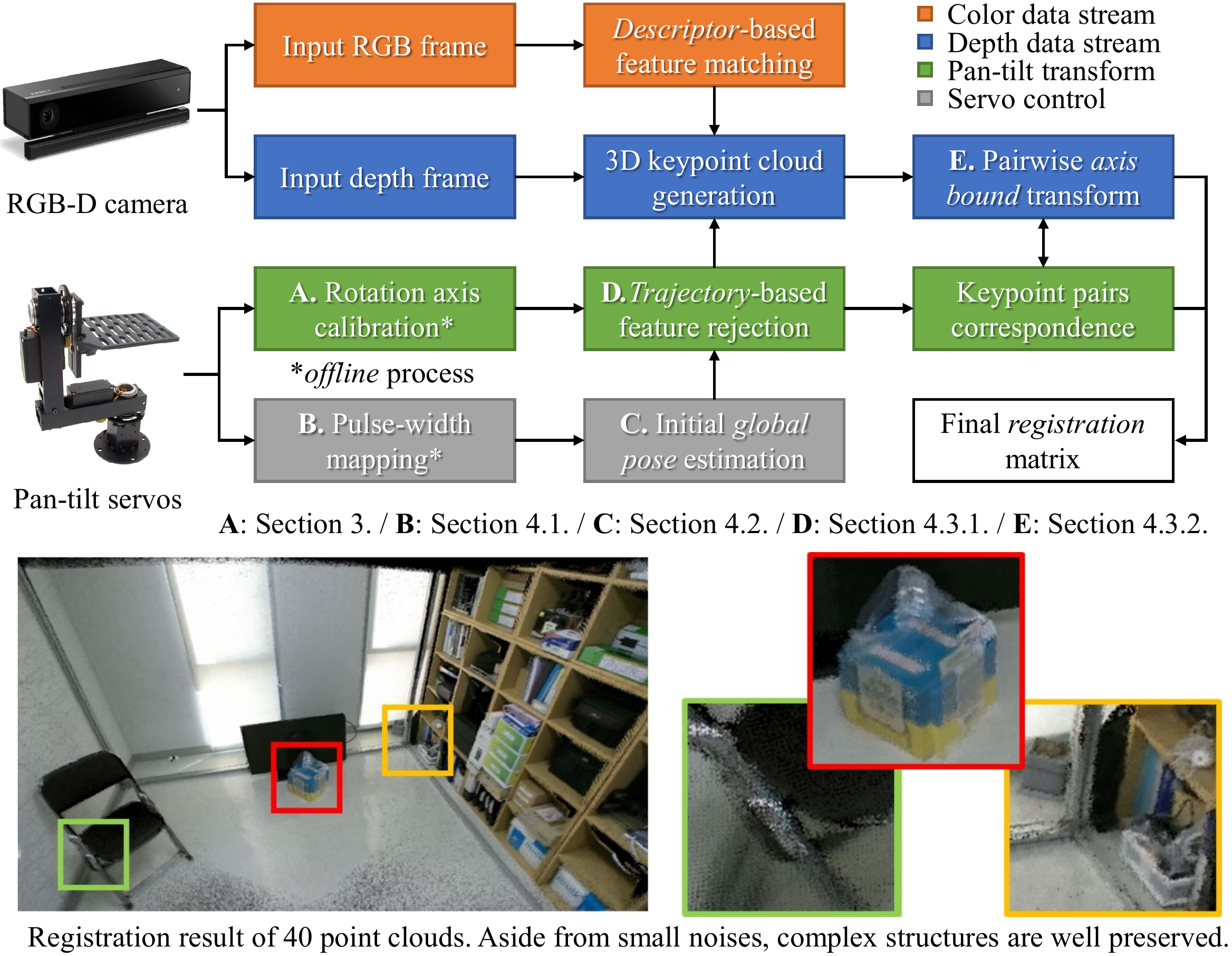}
\vspace{-1.\baselineskip}
\caption{\label{fig:representative_process}
The overall process and registration result of the proposed method.
}
\vspace{-1.\baselineskip}
\end{figure}

\begin{figure*}[!htb] 
\centering
\subfloat{
   \includegraphics[page=1, width=\linewidth]{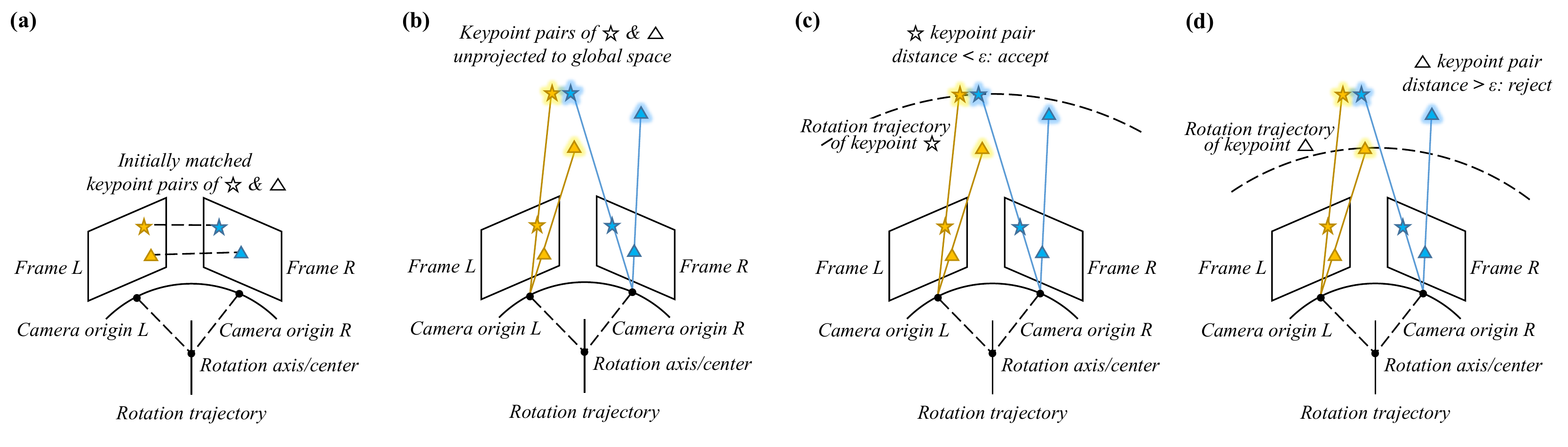}
}
\vspace{-1.\baselineskip}
\caption{\label{fig:keypoint_match_reject_process}
Schematic diagram for rejecting outlier matching pairs based on trajectory constraints.
For detailed descriptions, readers are referred to Section~\ref{sec:local_registration}.
(a) Keypoint pairs between two proximate frames are matched in the 2D image domain. 
(b) Keypoints are unprojected to 3D points in \emph{global} space, represented with halos.
(c) Keypoints whose distance is below the threshold $\epsilon$ are accepted as a correspondence pair.
(d) Keypoints whose distance is above the threshold $\epsilon$ are rejected as outlier matches.}
\vspace{-.75\baselineskip}
\end{figure*}

As SLAM methods can perform localization of the camera that is freely moving and mapping of the environment at the same time, some pan-tilt setups adopts SLAM methods for such purposes. For example, in \citet{wilson2012steerable}, KinectFusion \citep{newcombe2011kinectfusion} was used to reconstruct the surface geometry for steerable projection AR.
However, many SLAM methods are based on iterative closest point (ICP) or bundle adjustment (BA) methods, which are still computationally heavy for non-high-end computers.
In this study, a fast and accurate registration algorithm, namely \emph{axis bound registration}, is proposed to reconstruct indoor environments with pan-tilt RGB-D cameras.
The overall process of the proposed method is shown in Figure~\ref{fig:representative_process}.
When compared to other state-of-the-art registration and SLAM methods, the proposed algorithm is able to produce better registration results in much shorter time.
We summarize the contributions of the presented paper as follow:
\begin{enumerate} 
\item Rotation axis calibration and servo control scheme are incorporate into registration algorithm to roughly estimate the global pose of the camera (Figure~\ref{fig:pan_tilt_model}).
\item A novel rotation trajectory-distance constraint is developed to robustly reject outlier keypoints, and to further refine the estimated pose without RANSAC (Figure~\ref{fig:keypoint_match_reject_process}).
\item The alternating optimization linearizes and removes the iterative solving of the non-linear problem, which greatly improves registration accuracy and speed (Table~\ref{tab:registration_D815_table} and \ref{tab:registration_D816_table}).
\end{enumerate}

\section{Related Work} \label{sec:related_work}

\subsection{Steerable Platform Calibration}

Pan-tilt--zoom (PTZ) cameras have long been used for surveillance, object tracking and so forth.
\citet{davis2003calibrating} presented a calibration model where the typical pin-hole camera model was extended to incorporate the characteristics of pan and tilt motions. 
In \citet{wu2013keeping}, Wu and Radke proposed a camera model involving not only pan and tilt motions but also zoom on the image domain. 

A calibration method for a rotating turntable using external camera was introduced in \citet{chen2014rotation}. 
The method was later extended by \citet{niu2017calibration}, where the camera was attached to a rotating plate. 
The method was also incorporated in \citet{byun2018control}, which proposed a control mechanism for the motion and orientation of a generalized pan-tilt camera. 
Combining rotation axis calibration and camera-servo control, the study adopted the inverse kinematics approach to accurately interpret and manipulate  the camera motion.

\citet{tsai2017indoor} took an approach that is similar to the proposed method, in that the pan-tilt system was used to estimate the current pose of the camera.
Multi-view calibration was performed to construct the database of transformations at each preset pan-tilt rotation. 
Later, the transformations were looked up from the database to seed the ICP algorithm for point cloud registration.
However, the method is limited in terms of scalability, as the method cannot estimate the pose of the camera when the pan-tilt platform rotates to unseen positions.

\subsection{Point Cloud Matching}

Super 4-points congruent sets (Super4PCS) \citep{mellado2014super} and fast global registration (FGR) \citep{zhou2016fast} are examples of global registration methods, which globally search the point correspondences between two point clouds of any condition.
The authors of \citep{mellado2014super} proposed a novel method for removing the quadratic time complexity of its predecessor in \citep{aiger20084}. The key idea behind this improvement is the use of the data structure in solving the core instance problem, where the goal is to find all candidate pairs of a given point that are within a distance range.
FGR optimizes an objective function involving candidate matches over the surface of the object scans to align surfaces. The authors argue that the method does not require initialization, yet it can achieve accuracy comparable to that of well-initialized local refinement algorithms.
Both global registration methods are compared with the proposed method in Section~\ref{sec:experiment_and_evaluation}.

\subsection{Localization and Mapping}

Research in the SLAM field focuses on positional, orientational tracking of the camera and reconstruction of the scene in tandem. 
As mentioned above, some SLAM methods rely on the ICP algorithm \citep{besl1992method} to track the pose of the camera.
For example, KinectFusion \citep{newcombe2011kinectfusion} and \citet{endres20143} tracks the pose of the camera by repeatedly revising the transformation to minimize the difference between two clouds of points based on ICP. 

The iterative nature of the ICP method makes SLAM algorithms inherently heavyweight in terms of computing resources, such as power and memory consumption.
Thus, several ICP-based SLAM methods is able to perform in real-time only with the assistance of GPU \citep{neumann2011real}. 
\citet{newcombe2011kinectfusion}, \citet{endres20143} and \citet{Whelan16ijrr}, all make significant use of the GPU computation for mapping construction as well as tracking and pose estimation.

To tackle the heavy computation of ICP, ORB-SLAM2 \citep{mur2017orb} was proposed.
The algorithm adopted BA method instead of ICP, which was sufficiently lightweight to  perform sparse reconstruction with a standard CPU.
However, even ORB-SLAM2 relies on CPU multi-threading, OpenMP, to achieve real-time performance.
Without multi-threading, the performance degraded to 4 frames per second (Table~\ref{tab:registration_D815_table} and \ref{tab:registration_D816_table}).

\section{Pan-Tilt Rotation Calibration} \label{sec:pantilt_rotation_modeling}

\begin{figure}[!htb]
\centering
\subfloat[]{
\includegraphics[width=0.45\linewidth]{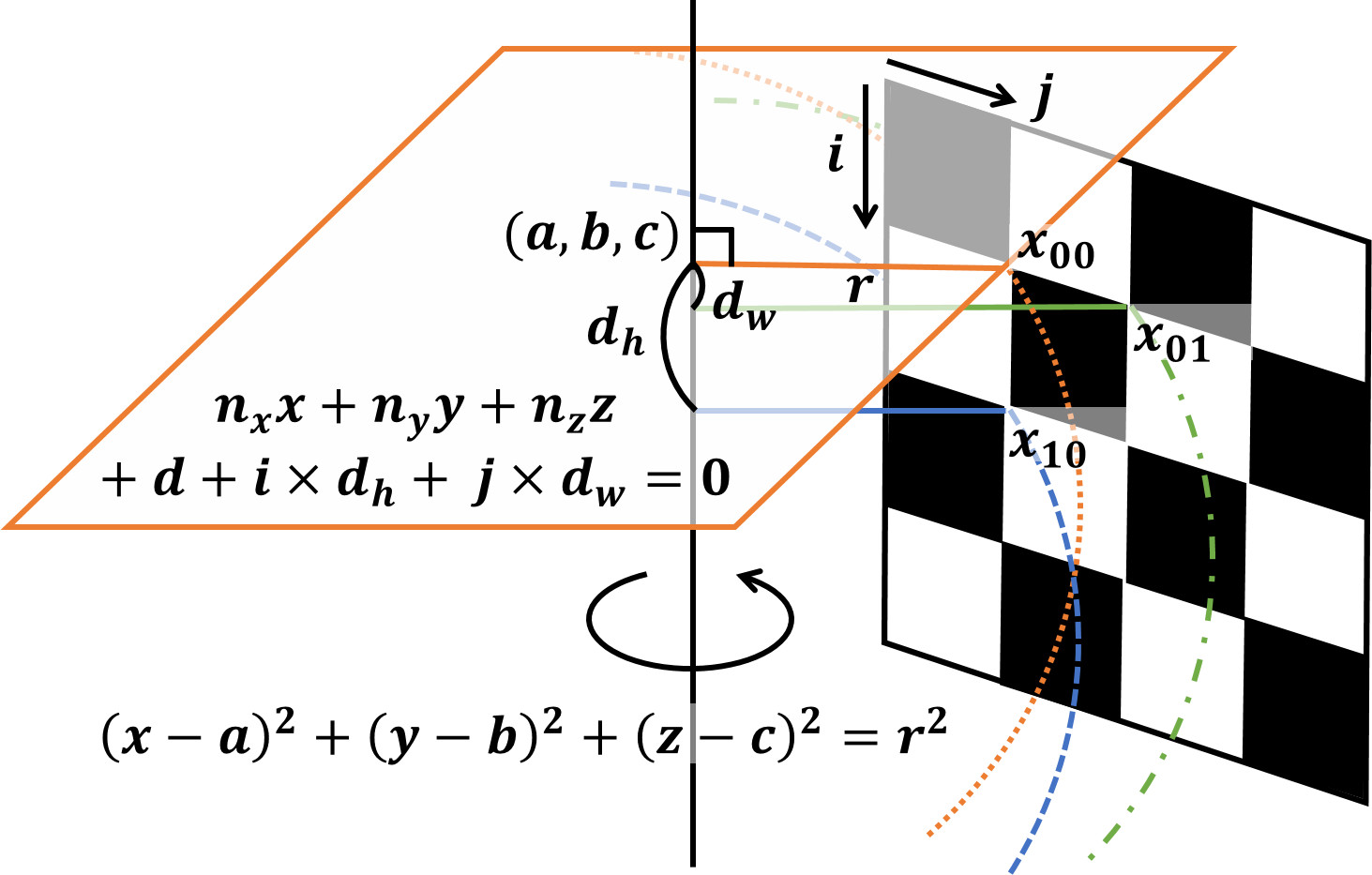}
}
\subfloat[]{
\includegraphics[width=0.5\linewidth]{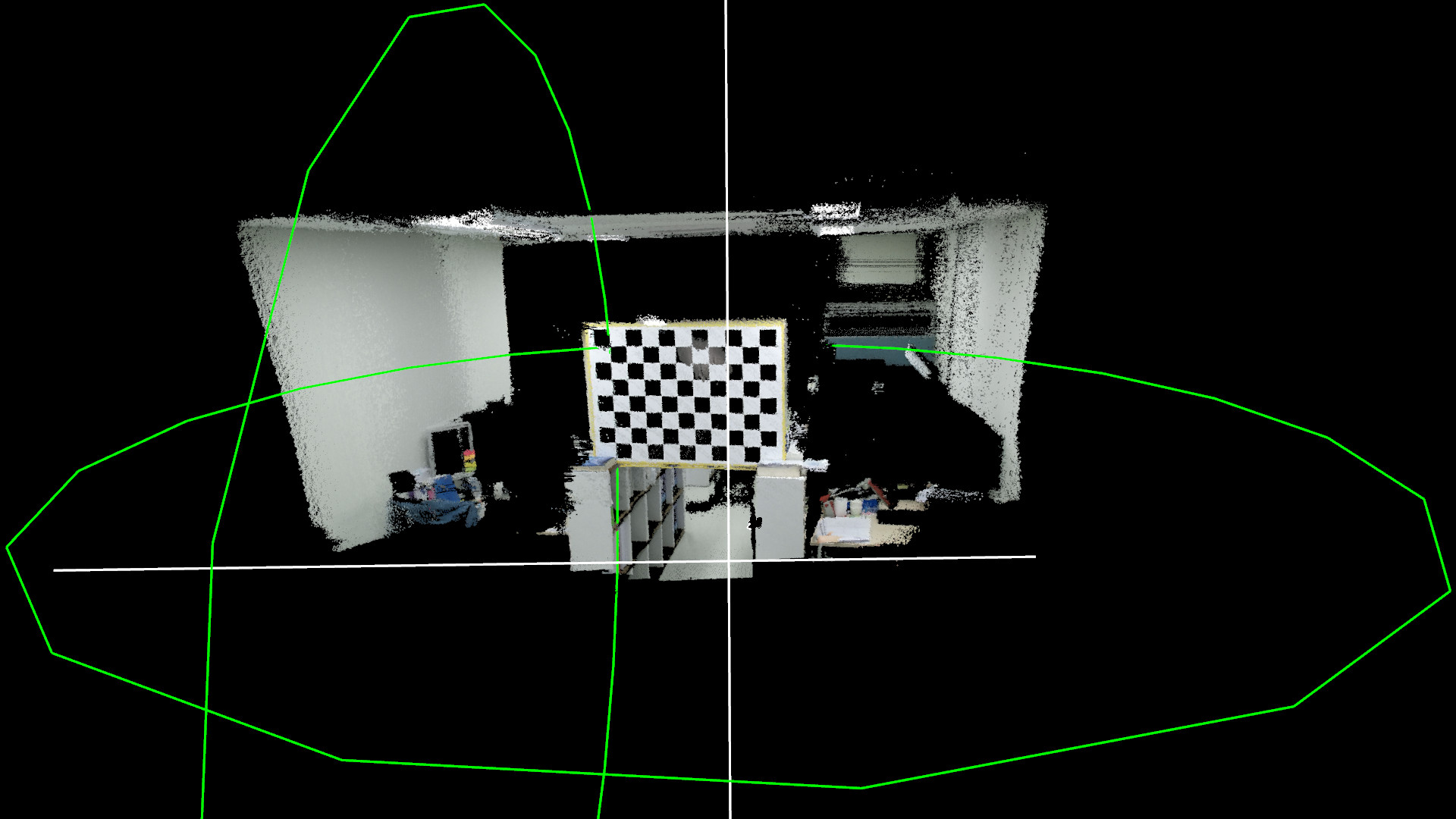}
}

\caption{\label{fig:pan_tilt_model}
(a) Rotation axis model \textcopyright\ \citet{byun2018control}.
(b) Pan-tilt calibration visualization. 
Green circles denote rotation trajectories and white lines denote their axes.
The rotation trajectories are calibrated with respect the top-leftmost corner, and thus exactly coincide at the corner.
}
\end{figure}

The proposed method takes into account the pan-tilt platform that controls the camera movement.
Thus, the rotation parameters of the pan-tilt servos are first calibrated, which are depicted in Figure~\ref{fig:pan_tilt_model}. 
The steps in \citet{byun2018control} are followed to recover their parameters for the rotation model.

Rotating points around a rotation axis forms closed circular trajectories, which is represented as great circles on 3-dimensional planes.
Thus, by fitting planes and circles the rotation direction vector and its center can be estimated.
The calibration starts by setting the initial frame as the \emph{reference} frame. 
Then, images of the checkerboard placed against the wall are captured while the camera rotates using pulse width modulation (PWM), from one end of the field-of-view to another.
Since the geometric relations of checkerboard corners are pre-known, all rotation trajectories can be represented relative to that of the top-leftmost corner in the \emph{reference} frame.
Finally, the aggregated corners from multi-view checkerboard frames are used to obtain 
rotation parameters of Figure~\ref{fig:pan_tilt_model} using all corner points with constrained global optimization \citep{chen2014rotation}.

Formally, the pan-tilt transform is modeled as follow. 
Let the rotation direction vector be defined as $n = [n_x,\ n_y,\ n_z]^\intercal$, with its norm being $||n|| = 1$, and the rotation center for the upper-left corner be defined as $p = [a,\ b,\ c]^\intercal$.
With all the parameters of $pan$ and $tilt$ rotations calibrated, the corresponding coordinates of 3D points between a \emph{local} camera frame and the \emph{reference} frame can now be represented.
If the $pan$ and $tilt$ angles of the \emph{local} frame are denoted by $\alpha$ and $\beta$ respectively, and the local point by $P_{local}$, its coordinate in the \emph{reference} frame can be modeled as follows:
\begin{equation}\label{eq:pantilt_model_global}
\begin{gathered}
\begin{bmatrix}
P_{\text{\emph{ref}}}  \\ 
1 
\end{bmatrix}
= T_{pan}\ R_{pan}(\alpha)\ T^{-1}_{pan}\ T_{tilt}\ R_{tilt}(\beta)\ T^{-1}_{tilt}\ 
\begin{bmatrix}
P_{local}  \\ 
1 
\end{bmatrix}
\end{gathered}
\end{equation}

Here, $R(\theta)$ is a $4\times 4$ matrix representing the rotation around a axis vector $n = [n_x\ n_y\ n_z]^\intercal$ and $T$ is a $4\times 4$ translation matrix of the position of the rotation axis $p = [a\ b\ c]^\intercal$.
Any point on the line of the rotation axis can be a pivot for the rotation, which does not affect the final transformation.
Here, the center of the rotation of the upper-leftmost corner of the checkerboard was used.
The same representation applies to both the $pan$ and $tilt$ transforms.

\section{Pan-Tilt Axis Bound Registration}

\subsection{Camera Transformation with Servo Control}\label{sec:transform_servo_control}

The proposed method roughly estimates the camera's pose based on the servo control. 
Potentiometers, which servos use to orient themselves, rotates linearly to the applied pulse width.
Therefore, it is possible to estimate the rotated angle of the pan-tilt platform based on the width of the applied pulse.
With rotation angles and applied pulse widths computed during calibration, a linear system can be constructed to estimate the rotated angle of the pan-tilt platform based on PWM as follows:
\begin{equation} \label{eq:pulse_angle_mapping}
[\hdots angle_{i} \hdots]^\intercal = scale \times [\hdots pulse_{i} \hdots]^\intercal + \text{\emph{offset}}.
\end{equation}

\begin{figure}[!htb]
\centering
\subfloat[Linear mapping between the pulse width and rotation angle.]{
\includegraphics[width=0.4\linewidth]{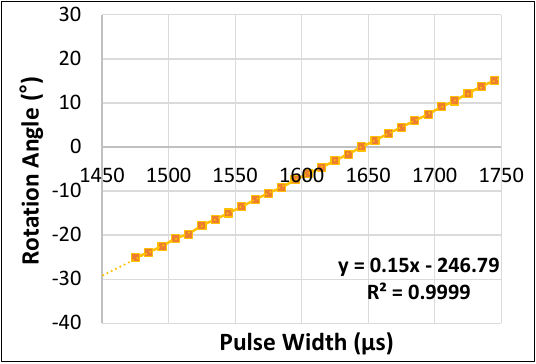}
}
~
\subfloat[Scatter plot of errors in the rotation angle estimation.]{
\includegraphics[width=0.4\linewidth]{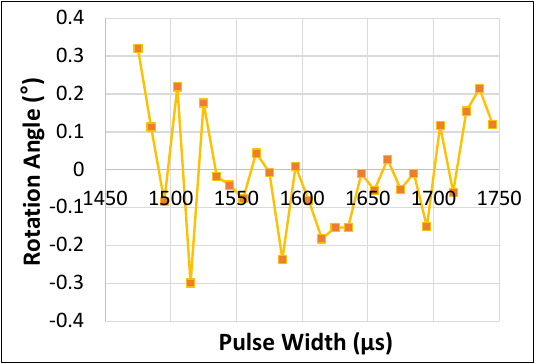}
}

\subfloat[The rotation estimation error histogram and its normal distribution.]{
\includegraphics[width=0.5\linewidth]{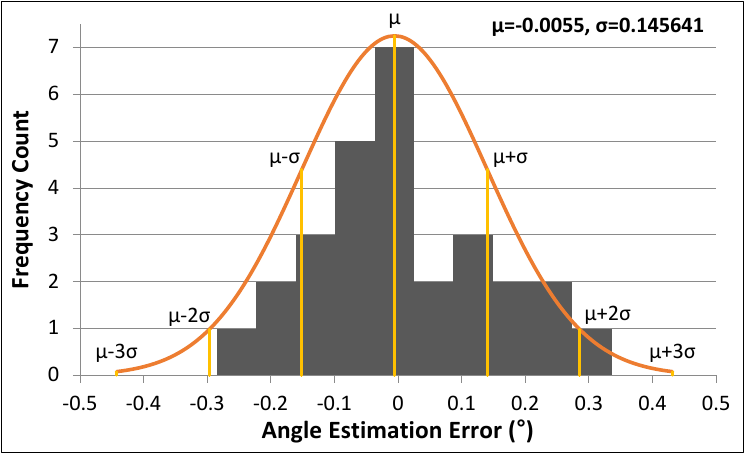}
}

\caption{\label{fig:pulse_angle_mapping}
Relationship plots for the pulse width and rotation angle.}
\vspace{-.5\baselineskip}
\end{figure}

Solutions for the above equation can be conveniently obtained by minimizing the sum of squared errors using Singular Value Decomposition (SVD).
Figure~\ref{fig:pulse_angle_mapping} (a) depicts the linear mapping result between the pulse widths and rotation angles.

We further refine the angle estimation, by modeling the residual of the estimation (Figure~\ref{fig:pulse_angle_mapping} (b)) as Gaussian noise (Figure~\ref{fig:pulse_angle_mapping} (c)).
The result of a Shapiro-Wilk test for normality \citep{shapiro1965analysis} strongly support that the residual distribution is normal, since p-value $>$ $\alpha$=.05 (W=.982, df=28 p=.898).

This is to compensate for two servo errors, namely the mechanical error and the random error, as categorized by \citet{wu2013keeping}.
The mechanical error refers to errors in angle estimation due to servo's manufacturing quality such as Figure~\ref{fig:pulse_angle_mapping} (b).
To compensate for mechanical errors, \citet{chao2014vision} added an additional error compensation term and used the spline curve for modeling such errors. 

\begin{figure}[!htb]
\centering
\subfloat[Image taken before the rotation.]{
\includegraphics[width=0.3\linewidth]{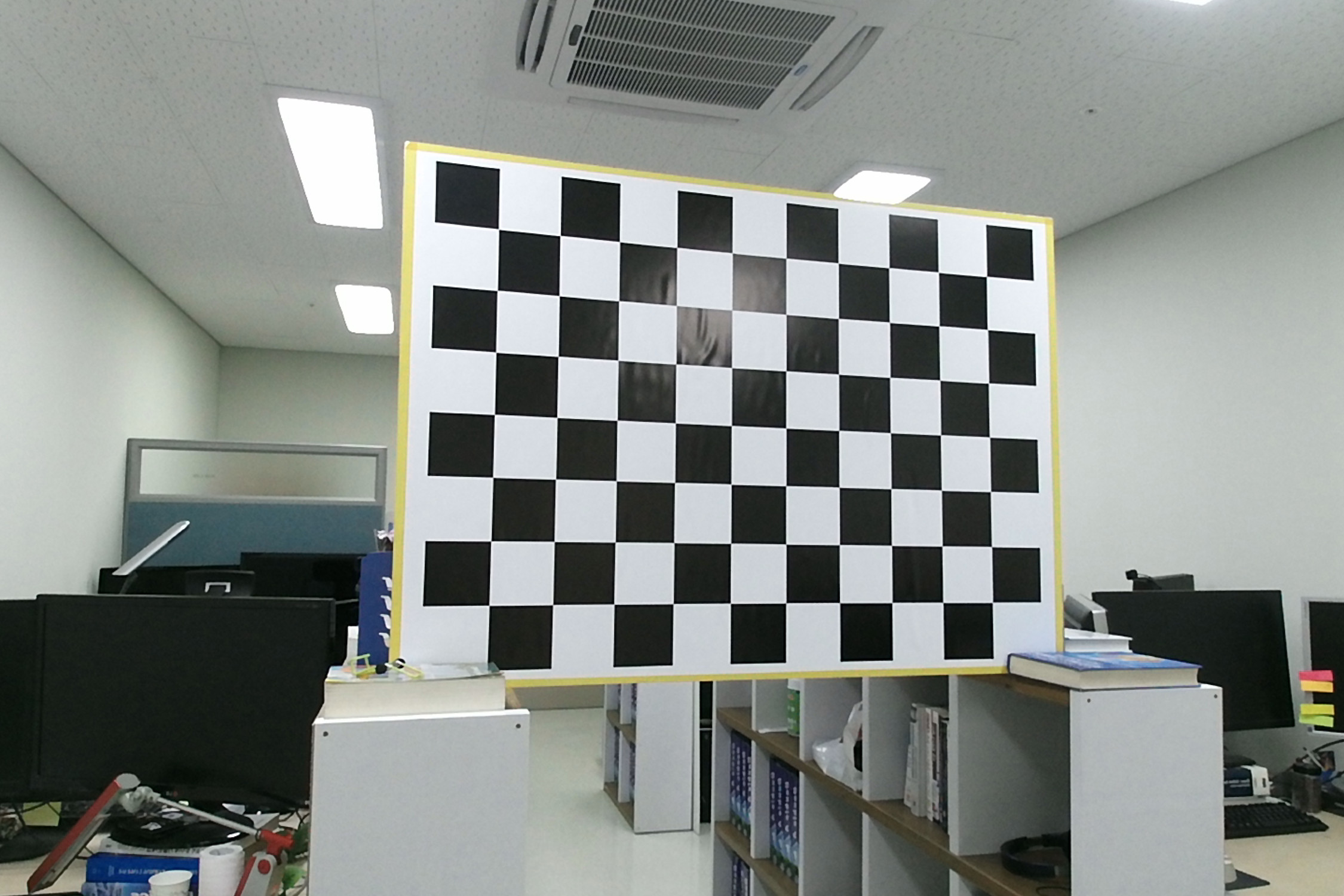}
}
~
\subfloat[Image taken after the rotation.]{
\includegraphics[width=0.3\linewidth]{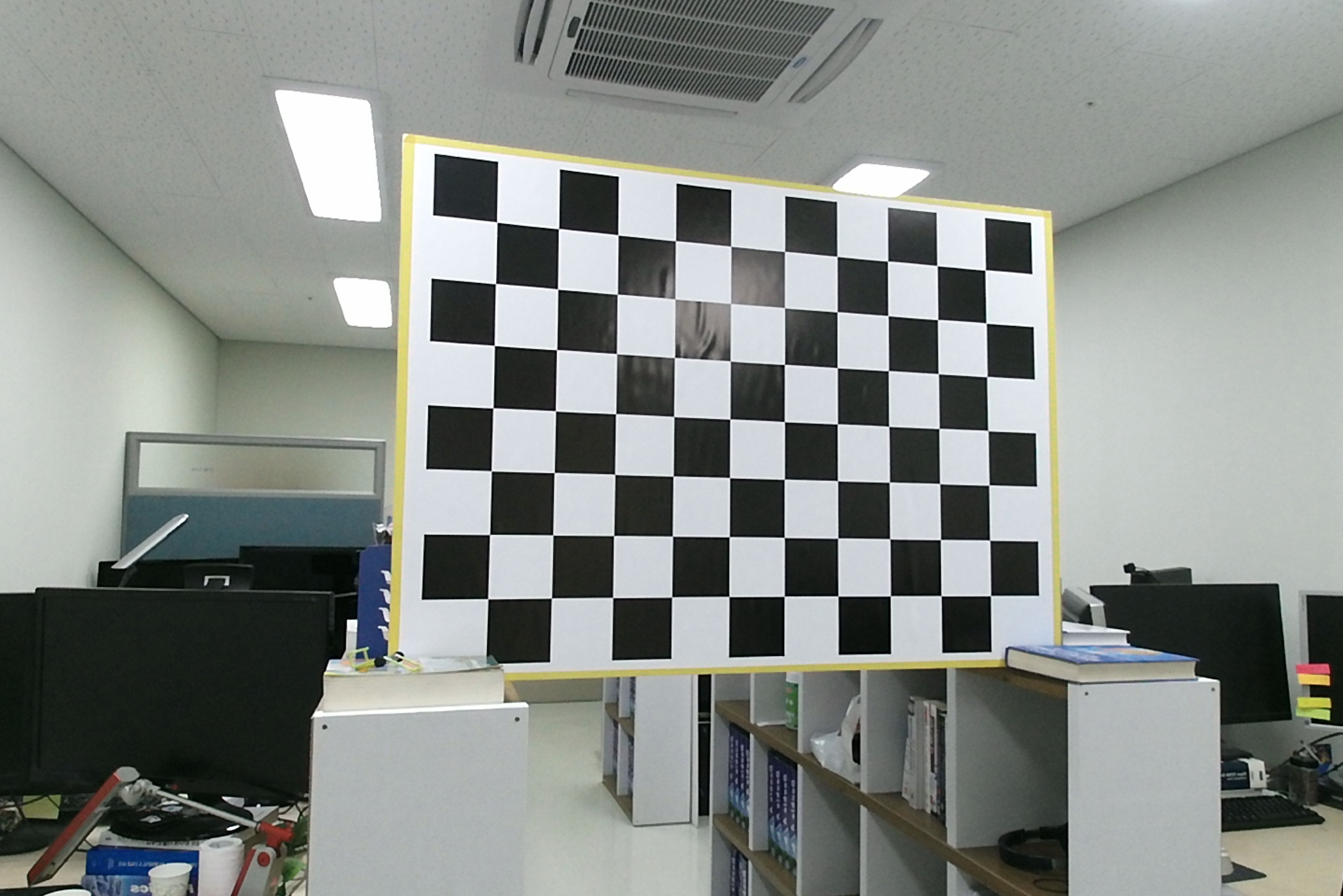}
}
~
\subfloat[Difference image of (a) and (b).]{
\includegraphics[width=0.3\linewidth]{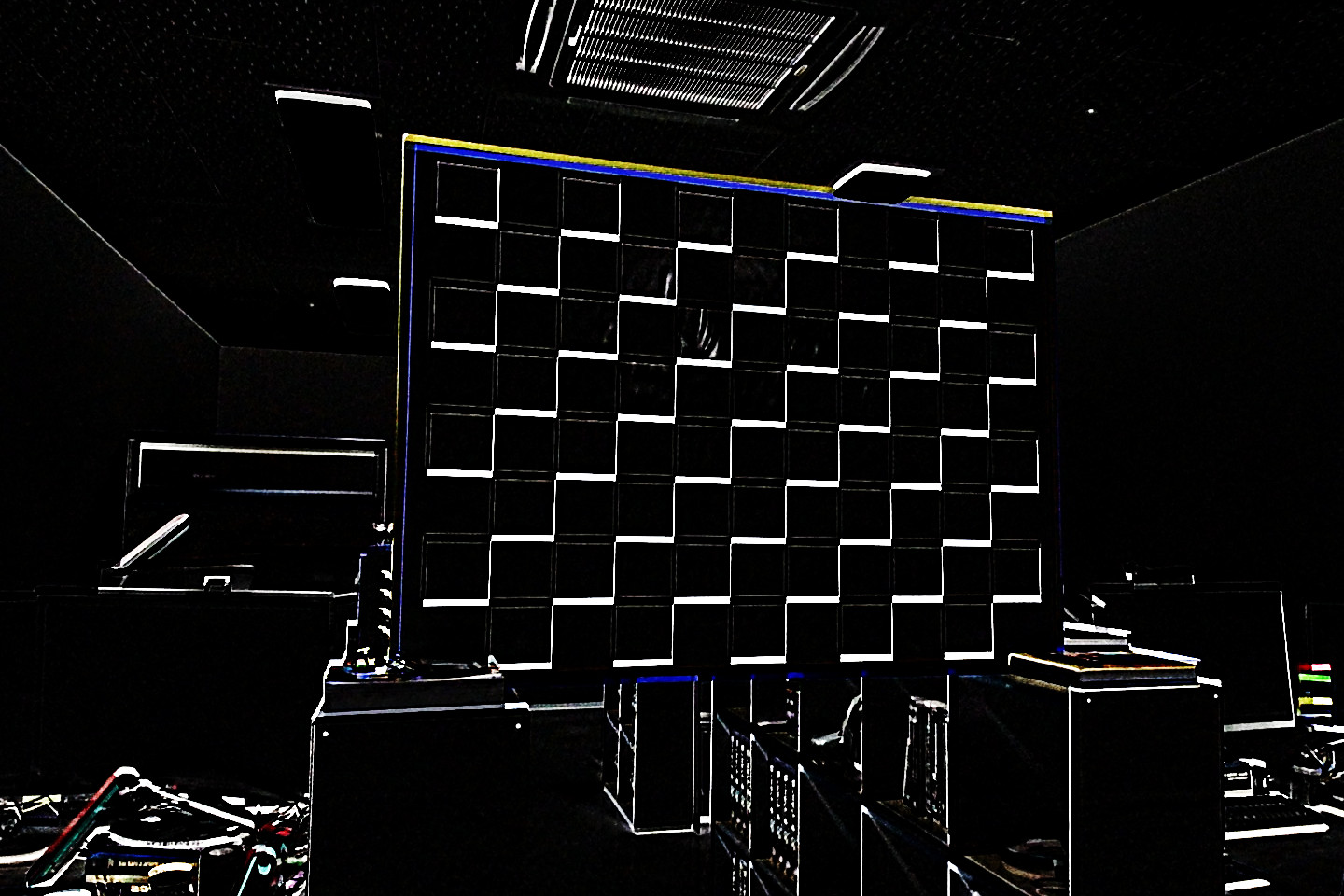}
}

\caption{\label{fig:pantilt_random_error}
Random errors of the pan-tilt system. Two 1920 $\times$ 1080 frames (a) and (b) were captured at the same servo position, controlled with  PWM. Thus, they should be identical, but in reality they differ by several pixels.}
\vspace{-.5\baselineskip}
\end{figure}

However, fitting raw errors to a fixed curve model cannot handle the other source of the error, which is the random error.
As demonstrated in Figure 4 of \citet{wu2013keeping}, a servo motor randomly deviates from its supposed position throughout continuous operation.
Such random errors with the pan-tilt system were also experienced in this study, as shown in Figure~\ref{fig:pantilt_random_error}.
Considering the randomness nature of the errors, we modeled the angular errors as Gaussian noise occurring to the linear mapping between the pulse width and the rotation angle.

\subsection{Global Registration} \label{sec:global_registration}

The rotation platform of the camera system is governed by two pan-tilt servos, which are controlled according to the width of the applied pulse.
Thus, by formulating the rotation angles and transformation model of the pan-tilt servos, the overall transformation of the camera system can be identified.
The rotation angle can be estimated by inputting the pulse width to Equation~\ref{eq:pulse_angle_mapping}, which is then substituted into Equation~\ref{eq:pantilt_model_global}, yielding the initial transform matrix.

The transform matrix is an estimation of the camera pose rotated by the pan-tilt servos.
Thus, the matrix can be used to register point clouds of multiple \emph{local} frames in the common \emph{global} frame.
However, as shown in Figure~\ref{fig:pulse_angle_mapping}, the presence of random errors in the pan-tilt servos indicates that the transformation obtained is still a coarse  estimation and should be refined further.
We compensate for the impact of such errors modeled as Gaussian noise in the \emph{local} registration.

\subsection{Local Registration} \label{sec:local_registration}

\subsubsection{Outlier Rejection with Rotation Trajectory Prior}

In pairwise local registration, feature matching between adjacent frames is performed.
However, simple matching of feature descriptors often suffers from false positives and outliers.
Although the random sample consensus (RANSAC) method is commonly employed to resolve this issue, the iterative nature of random sampling inevitably requires expensive computation, but still cannot guarantee the correct result.
Thus, we propose a rejection method for outlier matches by leveraging the prior knowledge of  the rotation trajectory obtained in Section~\ref{sec:transform_servo_control}.

Let the first and second frames of a given pair be denoted by $l$ and $r$, respectively. 
A frame consists of a color image and corresponding depth data.
To start with, we establish a set $\mathcal{K}_{lr}$ of putative correspondences by extracting keypoints and matching their descriptors from color images, using ORB \citep{rublee2011orb}.
Using the depth data, the keypoint pairs in the image domain are then projected to each \emph{local} camera frame.
3D points of the $i$-th keypoint pair in the set are denoted by $k^{l}_{i}$ and $k^{r}_{i}$.
From Equation~\ref{eq:pulse_angle_mapping}, \emph{pan} and \emph{tilt} angles of each frame can be acquired, namely $\alpha^{l}$, $\beta^{l}$, $\alpha^{r}$ and $\beta^{r}$.
Substituting $k_{i}$, $\alpha$, $\beta$ into Equation~\ref{eq:pantilt_model_global} yields pre-oriented 3D points in the \emph{global} frame, denoted by $\hat{k}_{i}$ as follows:
\begin{equation}\label{eq:global_transformed_keypoints}
\begin{gathered}
\begin{bmatrix}
\hat{k}^{l}_{i}  \\ 
1 
\end{bmatrix}
= T_{pan}\ R_{pan}(\alpha^{l})\ T^{-1}_{pan}\ T_{tilt}\ R_{tilt}(\beta^{l})\ T^{-1}_{tilt}\ 
\begin{bmatrix}
k^{l}_{i} \\ 
1 
\end{bmatrix}, 
\\
\begin{bmatrix}
\hat{k}^{r}_{i}  \\ 
1 
\end{bmatrix}
= T_{pan}\ R_{pan}(\alpha^{r})\ T^{-1}_{pan}\ T_{tilt}\ R_{tilt}(\beta^{r})\ T^{-1}_{tilt}\ 
\begin{bmatrix}
k^{r}_{i} \\ 
1 
\end{bmatrix}.
\end{gathered}
\end{equation}

Given pre-oriented keypoint pairs $\hat{k}^{l}_{i}$ and $\hat{k}^{r}_{i}$, we determine whether they are falsely matched or not based on their distances to the rotation trajectory, as illustrated in Figure~\ref{fig:keypoint_match_reject_process}.
Specifically, the indicator function $g(\hat{k}^{l}_{i},\ \hat{k}^{r}_{i})$ declares that the matching of $\hat{k}^{l}_{i}$ and $\hat{k}^{r}_{i}$ is a false positive if the distance exceeds a certain limit, namely $\epsilon$, as follows:
\begin{equation}\label{eq:distance_threshold_rejection}
  g(\hat{k}^{l}_{i},\ \hat{k}^{r}_{i})=\left\{
  \begin{array}{@{}ll@{}}
	1, & \text{if}\ \|\hat{k}^{l}_{i} - \hat{k}^{r}_{i}\|_{2} < \epsilon \\
	0, & \text{otherwise}
  \end{array}\right.
\end{equation}

The value of $\epsilon$ is dynamically imposed for each point. 
We first consider that the estimated rotation angles (Equation~\ref{eq:pulse_angle_mapping}) exhibit Gaussian-distributed errors.
Given the standard deviation $\sigma$, we allow $\pm3\sigma$ range of deviation for the estimated rotation angle.
As Pr$(\mu-3\sigma\leq X\leq\mu+3\sigma)\approx0.9973$ for a normally distributed random variable, we assume that estimations that deviate more than the $\pm3\sigma$ slack are outliers.
Since the camera rotates around the circular trajectory, the distance between rotated points can be approximated as the arc length.
Thus, for a true-positive match, $\|\hat{k}^{l}_{i} - \hat{k}^{r}_{i}\|_{2} < \epsilon\approx d\theta$, where $d$ is the distance of a point from the camera and $\theta$ is the maximal allowed angular deviation in radian.
In the proposed setup, $d$ is set as $\|\hat{k}^{l}_{i}\|_{2}$, and $\theta$ as $3\sigma\pi/180$, considering the $\sigma$ value in Figure~\ref{fig:pulse_angle_mapping}.

\subsubsection{Pairwise Transform with Axis Bound Registration} \label{sec:axis_bound_registration}

In pairwise registration, the objective of local registration is to determine the pan-tilt angles $\alpha^{r}$ and $\beta^{r}$ of the frame  $r$, seeded from Equation~\ref{eq:pulse_angle_mapping}, that transforms $k^{r}_{i}$ to its correspondence $\hat{k}^{l}_{i}$ of the frame $l$, while $\alpha^{l}$ and $\beta^{l}$ are fixed.
Note that $k^{r}_{i}$ is a \emph{local} point, whereas $\hat{k}^{l}_{i}$ is pre-oriented as in Equation~\ref{eq:global_transformed_keypoints}.
Formally, the cost function for the local registration is constructed to minimize the distances between correspondence pairs as follow:
\begin{equation}\label{eq:transform_cost_function}
\resizebox{\hsize}{!}{$
\begin{aligned}
\arg\min_{\alpha^{r},\ \beta^{r}}\sum_{i\in\mathcal{X}}^{}
\Bigg\Vert 
\begin{bmatrix}
\hat{k}^{l}_{i}  \\ 
1 
\end{bmatrix} 
- T_{pan}\ R_{pan}(\alpha^{r})\ T^{-1}_{pan}\ T_{tilt}\ R_{tilt}(\beta^{r})\ T^{-1}_{tilt}
\begin{bmatrix}
k^{r}_{i} \\ 
1 
\end{bmatrix} 
\Bigg\Vert^{2}_{2},\\
\text{where } \mathcal{X} = \{i \mid (k^{l}_{i}, k^{r}_{i})   \in \mathcal{K}_{lr} \text{ and } g(\hat{k}^{l}_{i}, \hat{k}^{r}_{i})=1\},\ |\mathcal{X}|=N.\phantom{b=\,}
\end{aligned}
$}
\end{equation}

The transformation regarding the cost function is bound by the two rotation axes, $pan$ and $tilt$, hence the name \emph {Axis Bound Registration} of the proposed algorithm.
The cost function (Equation~\ref{eq:transform_cost_function}) is a non-linear optimization problem about two trigonometric variables $\alpha^{r}$ and $\beta^{r}$. 
Though the Levenberg-Marquardt method is commonly employed to solve the non-linear problem, iterative methods often hinder real-time performance.
Thus, we instead adopt alternating optimization partially linearize and accelerate the problem solver. 

Let $\theta_{tilt}$ denote the given value for the \emph{tilt} rotation. Then by the rotation transformation of Equation~\ref{eq:pantilt_model_global}, the cost function Equation~\ref{eq:transform_cost_function} can now be divided into the function for the $pan$ rotation angle $\alpha$ as follows:
\begin{equation}\label{eq:cost_function_alpha}
\begin{gathered}
\begin{aligned}
\arg\min_{\alpha}\sum_{i}^{N}
\Bigg\Vert 
\begin{bmatrix}
\hat{k}^{l}_{i}  \\ 
1 
\end{bmatrix}
- T_{pan}\ R_{pan}(\alpha)\ T^{-1}_{pan}\ 
\begin{bmatrix}
k^{r\prime}_{i}  \\ 
1 
\end{bmatrix}
\Bigg\Vert^{2}_{2}
\end{aligned}
.
\end{gathered}
\end{equation}
Similarly, $\theta_{pan}$ denotes the given value for the \emph{pan} rotation and the sub-function for the $tilt$ rotation angle $\beta$ is constructed as follows:
\begin{equation}\label{eq:cost_function_beta}
\begin{gathered}
\begin{aligned}
\arg\min_{\beta}\sum_{i}^{N}
\Bigg\Vert 
\begin{bmatrix}
\hat{k}^{l\prime}_{i} \\ 
1 
\end{bmatrix}
- T_{tilt}\ R_{tilt}(\beta)\ T^{-1}_{tilt}\ 
\begin{bmatrix}
k^{r}_{i}  \\ 
1 
\end{bmatrix}
\Bigg\Vert^{2}_{2}
\end{aligned}
.
\end{gathered}
\end{equation}
The $r$ superscripts of $\alpha$ and $\beta$ are omitted for visibility.
$\hat{k}^{l}_{i},\ \hat{k}^{l\prime}_{i},\ k^{r}_{i},\ k^{r\prime}_{i}$ are all 1$\times$3 column vectors denoting $XYZ$ coordinates in the 3D space. 
$[k^{r\prime}_{i}\ 1]^\intercal = T_{tilt}\ R_{tilt}(\theta_{tilt})\ T^{-1}_{tilt} [k^{r}_{i}\ 1]^\intercal$
and
$[\hat{k}^{l\prime}_{i}\ 1]^\intercal = T_{pan}^{-1}\ R_{pan}^{-1}(\theta_{pan})\ T_{pan} [\hat{k}^{l}_{i}\ 1]^\intercal$
are introduced to respectively incorporate known rotations of $\theta_{tilt}$ and $\theta_{pan}$. 

Each sub-cost function can be iteratively solved in closed form, in each alternating step.
However, albeit empirically in the experiment (Section~\ref{sec:experiment_and_evaluation}), we found that a single alternation sufficed to yield numerically optimal solutions.

To solve for Equations~\ref{eq:cost_function_alpha} and \ref{eq:cost_function_beta}, we rearrange all terms of $||[\hat{k_i}\ 1]^{\intercal} - TR(\theta_{(\cdot)})T^{-1}[k_i\ 1]^{\intercal}||$, $1\leq i\leq N$, into matrix expressions in the form of $\mathbf{A(\theta_{(\cdot)})x=b}$ to construct two linear systems, where 
$\mathbf{x} = [\cos(\alpha)\ \sin(\alpha)\ 1]^\intercal$ and 
$\mathbf{b} = [\hdots \hat{k}^{l}_{i} \hdots]^\intercal$ for $pan$,
and
$\mathbf{x} = [\cos(\beta)\ \sin(\beta)\ 1]^\intercal$ and 
$\mathbf{b} = [\hdots \hat{k}^{l\prime}_{i} \hdots]^\intercal$ for $tilt$.
Due to limited space, exact formulas for $\mathbf{A(\theta_{(\cdot)})}$ are shown in the supplementary materials.

Both $\mathbf{A(\theta_{pan})}$ and $\mathbf{A(\theta_{tilt})}$ can be solved by the SVD method, which yields least squares error solutions for $\cos(\beta),\ \sin(\beta)$, and $\cos(\alpha),\ \sin(\alpha)$.
The initial solutions are refined to obtain the actual rotation angles of interest by enforcing trigonometric properties, that is $\beta = \arctan2(\sin(\beta),\ \cos(\beta))$ and $\alpha = \arctan2(\sin(\alpha),\ \cos(\alpha))$.
The $\alpha$ and $\beta$ angles define the pairwise transformation from frame $r$ to $l$ in $global$ space, which finalizes $local$ registration between two frames.

\section{Experiment and Evaluation} \label{sec:experiment_and_evaluation}

\begin{figure}[!htb] 
\centering

\subfloat[Pan-tilt camera system.]{
\includegraphics[width=0.23\linewidth]{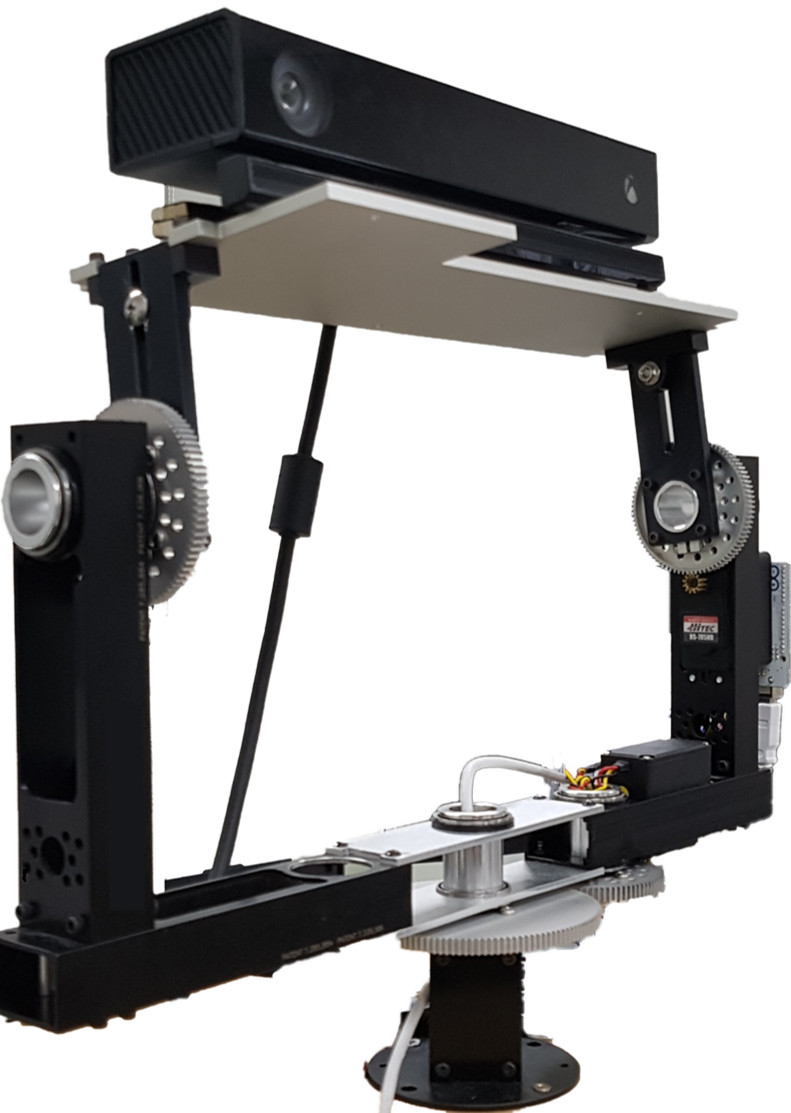}
}
~
\subfloat[Pan-tilt camera system fixed to the ceiling.]{
\includegraphics[width=0.5\linewidth]{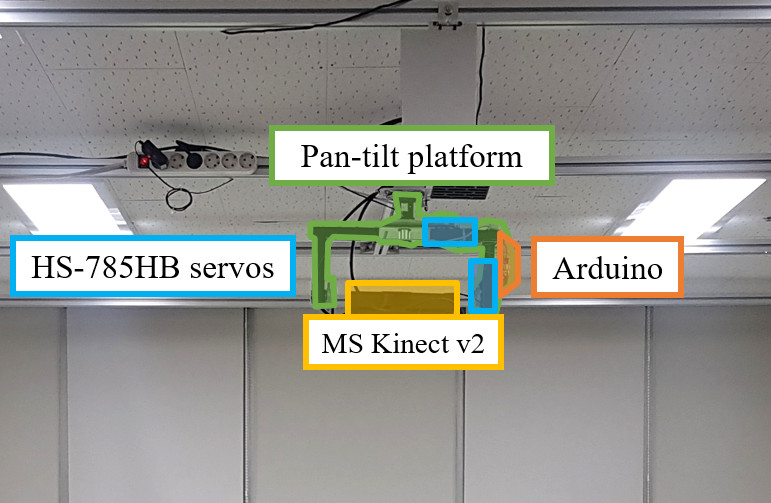}
}
\caption{\label{fig:pantilt_experiment_configuration}
Pan-tilt RGB-D scan registration configuration.}
\vspace{-.5\baselineskip}
\end{figure}

\subsection{System and Dataset Configuration}

To validate the proposed method, an experimental environment was set up comprising an RGB-D camera and two pan-tilt servos, supported by a steerable platform as shown in Figure~\ref{fig:pantilt_experiment_configuration}.
The steerable platform was assembled from custom-made arms and gears, where two servos were attached to rotate the platform in \emph{pan} and \emph{tilt} directions.
Two HS-785HB servo motors were used, which accept $600$--$2400$ microseconds pulse width. 
Microsoft Kinect v2 was used, for $1920\times 1080$ RGB and $512\times 424$ time-of-flight depth images. 

\begin{figure*}[!htb] 
\centering
\includegraphics[width=\linewidth]{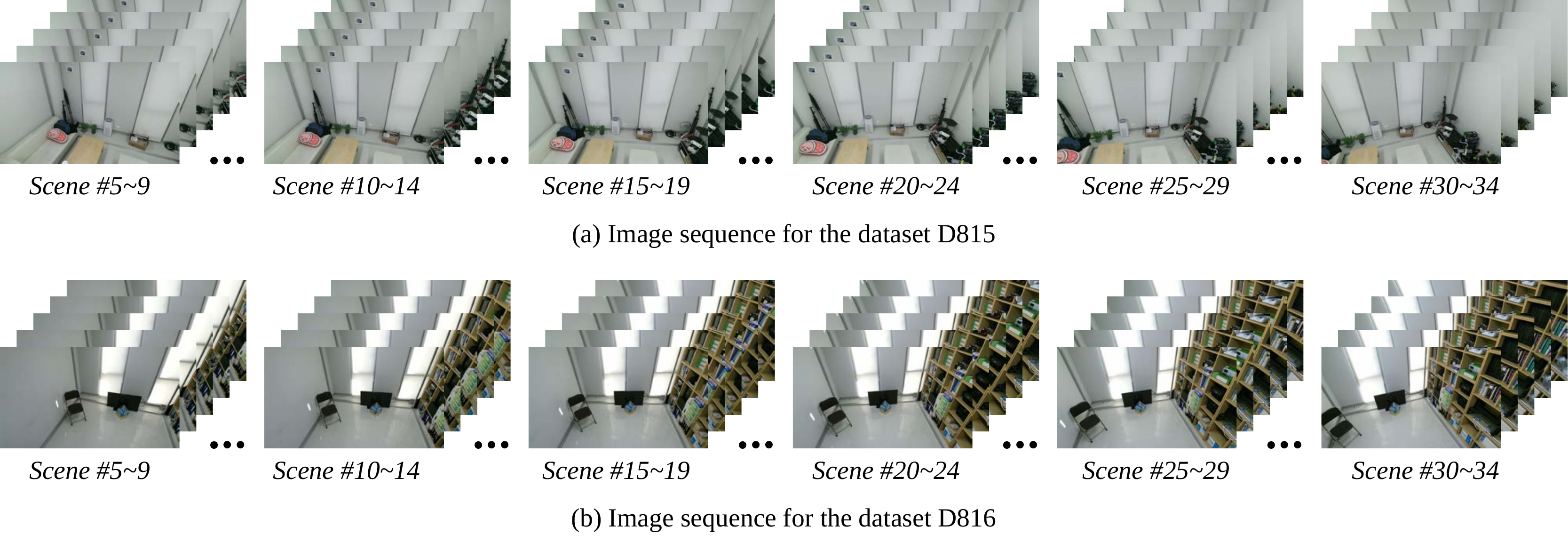}
\vspace{-1.8\baselineskip}
\caption{ \label{fig:register_input_scenes}
Two datasets, (a) Room D815 and (b) Room D816, used for pan-tilt RGB-D scan registration, consisting of 30 frames each.
Here, only RGB images are shown. In the experiment, depth data and point clouds were also provided.
}
\vspace{-.5\baselineskip}
\end{figure*}

Since the proposed approach combines servo control and camera pose estimation for point cloud registration, calibration data, including rotation axes and pulse width mapping, are required.
However, to the best of the authors' knowledge, there is no publicly available dataset that meets all the requirements for pan-tilt RGB-D scan registration, including the dataset of \citet{tsai2017indoor}, which used a similar pan-tilt camera setup.
Therefore, the pan-tilt RGB-D scan dataset was constructed in-house, the sequences of which are shown in Figure~\ref{fig:register_input_scenes}. 

\subsection{Pan-Tilt RGB-D Scan Registration}

The proposed method was compared with four state-of-art methods that used different approaches. 
The first is the RGBD-Calib \citep{tsai2017indoor}, which makes use of external parameters of the camera for pose estimation and used ICP algorithm to refine the poses.
Two other methods are FGR \citep{zhou2016fast} and S4PCS \citep{mellado2014super}, which were compared with the RGBD-Calib method in \citet{tsai2017indoor}. 
Both methods are global registration methods, which register two point clouds by searching for global point correspondences.
The last method for comparison is ORB-SLAM2 \citep{mur2017orb}.
ORB-SLAM2 is one of the most prominent SLAM methods, which performs pose estimation of a RGB-D camera, as well as a sparse 3D reconstruction.
Though ORB-SLAM2 supports CPU multi-threading, multi-threading option was disabled in the experiments for fairness as all other methods run in a single process/thread.

To quantitatively measure the performance of the registration results, the RMS metric of $N$ closest points was used as in \citet{tsai2017indoor}. Given $N$ points in the input point cloud, or frame $r$, their closest points are determined in the source point cloud, or frame $l$. If the vector storing the vector distances (in millimeters) between $N$ closest point pairs is denoted by $d_{min}\in R^N$ and its $i$-th element by $d_{min}(i)$, the RMSE is defined as 
\begin{equation} \label{eq:rms}
\text{RMSE} = \sqrt{\frac{\sum_{i=1}^{N} d^2_{min}(i)}{N} }.
\end{equation}
All experiments were conducted on a Windows 10 with Intel Core i7-6700K CPU @ 4.00 GHz and 16 GB of DDR4 memory. For image processing and camera calibration, the OpenCV library \citep{opencv_library} was used. For ICP algorithm and $N$ closest points RMSE implementation, the PCL library \citep{pcl_library} was used. All codes were written in C++ and compiled with the O2 optimization and no multi-threading.

\subsection{Experiment, Evaluation, and Analysis}

\begin{table*}[!htb] 
\centering
\caption{\label{tab:registration_D815_table}
The result tables for the experiment D815 in terms of (a) error and (b) speed.
RGBD-Calib \citep{tsai2017indoor}, ORB-SLAM2 \citep{mur2017orb}, FGR \citep{zhou2016fast}, and S4PCS \citep{mellado2014super} are compared with the proposed method.
}
\vspace{-.5\baselineskip}
\subfloat[RMSE results of the dataset D815. The lowest RMSEs are highlighted in green.]{
\includegraphics[width=.5\linewidth]{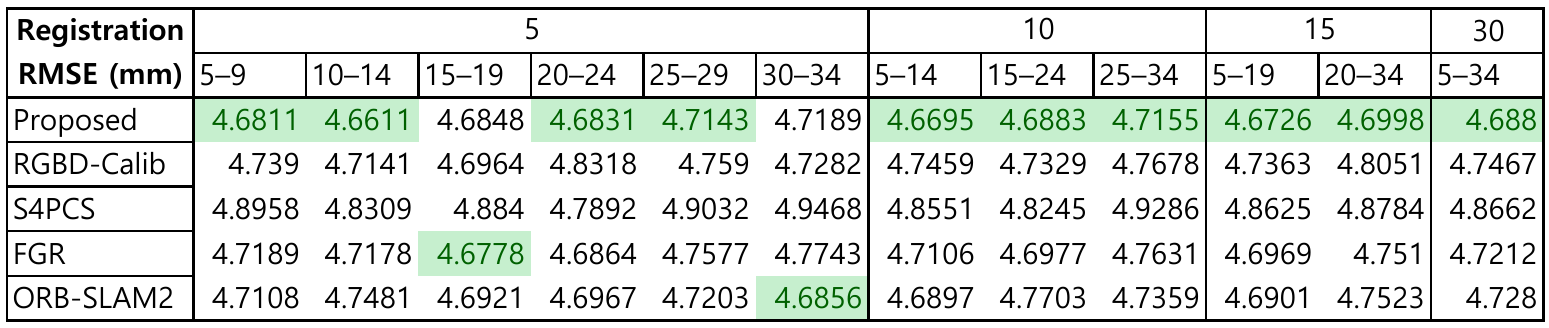}
}
~
\subfloat[Execution time of the dataset D815. The fastest times are highlighted in yellow.]{
\includegraphics[width=.5\linewidth]{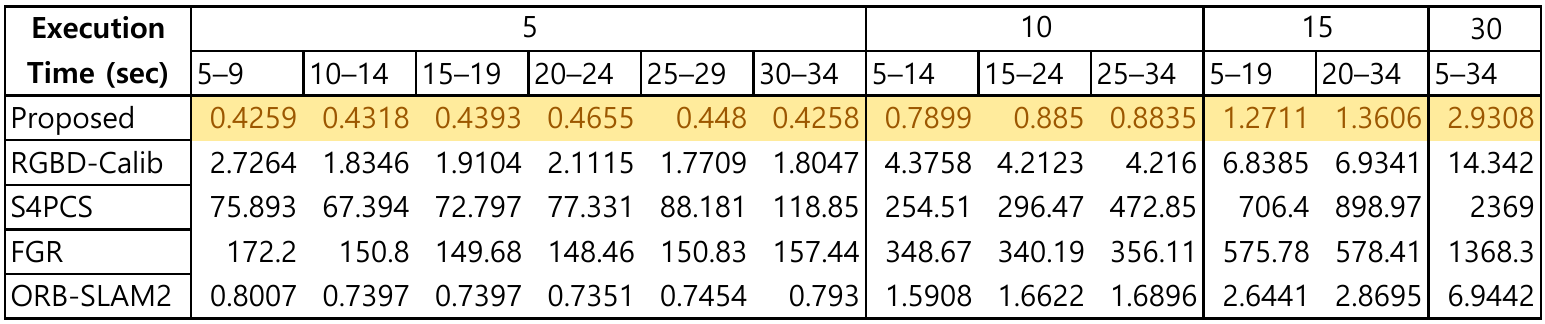}
}
\vspace{-.5\baselineskip}
\end{table*}

Figure~\ref{fig:register_input_scenes} shows two datasets for the registration experiments.
Named after the room number, D815 and D816, the datasets consist of 30 frames each, created by by rotating the camera in \emph{pan} direction by 1.5\textdegree\ at each step, with arbitrary \emph{tilt} angles.
To make the most of the limited number of scan data, the experiment was conducted with 6 partitions of the dataset, consisting of \emph{5 frames, 10 frames, 15 frames} and \emph{30 frames}, yielding total 12 test cases for each dataset.

For the quantitative evaluation, the registration results in terms of registration error and execution time are summarized in Table~\ref{tab:registration_D815_table} (D815), and in Table~\ref{tab:registration_D816_table} (D816).
For the visual comparison, the registration images of the experiment are available as Figure~A.1--A.12 (D815) and Figure~B.1--B.12 (D816) in the supplementary material.
We emphasize that a fair amount of time was devoted to adjusting parameters for each algorithm, particularly for FGR and S4PCS, to optimize the algorithm in terms of RMSE. 
Moreover, the termination conditions of ICP for RGBD-Calib were also optimized.

\subsubsection{Experiment on Dataset D815}

\begin{table*}[!htb] 
\centering
\caption{\label{tab:registration_D816_table}
The result tables for the experiment D816 in terms of (a) error and (b) speed.
RGBD-Calib \citep{tsai2017indoor}, ORB-SLAM2 \citep{mur2017orb}, FGR \citep{zhou2016fast}, and S4PCS \citep{mellado2014super} are compared with the proposed method.
}
\vspace{-.5\baselineskip}
\subfloat[RMSE of the dataset D816. The lowest RMSEs are highlighted in green.]{
\includegraphics[width=.5\linewidth]{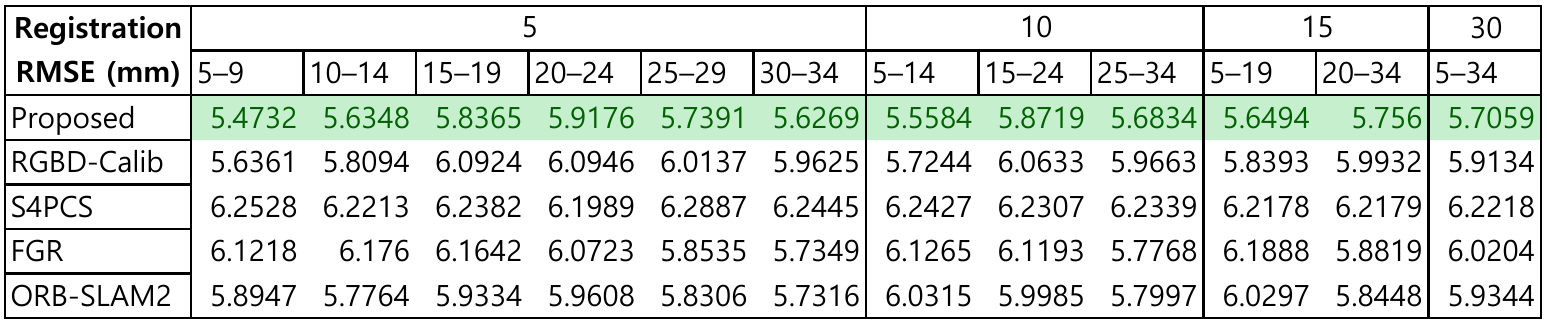}
}
~
\subfloat[Execution time of the dataset D816. The fastest times are highlighted in yellow.]{
\includegraphics[width=.5\linewidth]{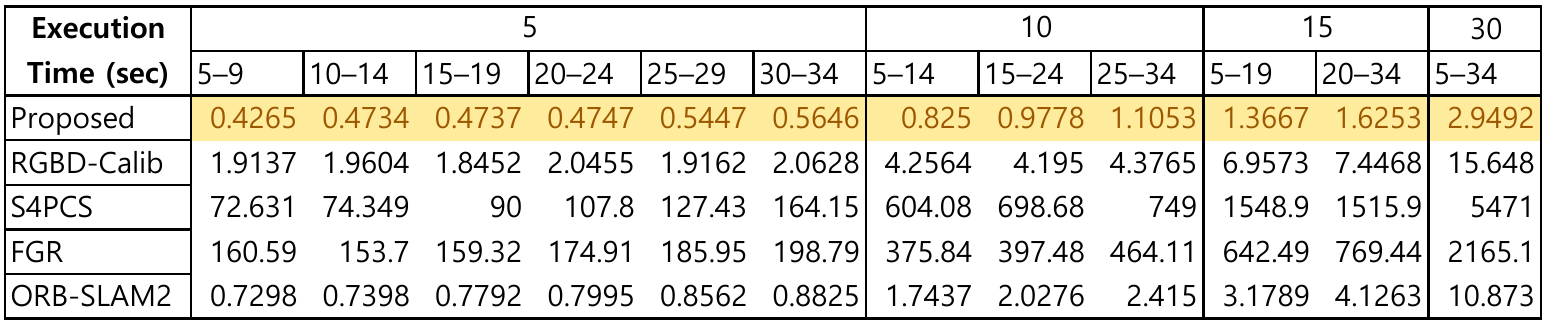}
}
\vspace{-.5\baselineskip}
\end{table*}

In terms of registration error, the proposed algorithm outperformed other algorithms in 19 cases out of all 21 test cases, and ranked second in the other 2 cases.
In terms of speed, the proposed algorithm was the apparently fastest in all 21 test cases, followed by the second fastest ORB-SLAM2.

It is notable that in many cases, the proposed method was up to more than twice as fast as the second ranked ORB-SLAM2, which is a BA-based SLAM method, while maintaining better performance in all but one test cases.
Also, the proposed method always outperformed RGBD-Calib, which is ICP-based calibration method, FGR and S4PCS, which are global registration methods,  in both error and speed terms.
The outperformance of the proposed method indicates that a fast and accurate registration is possible without ICP or BA method, by directly modeling 3D motion trajectories of the actuated camera, and utilizing those information in the registration.

\subsubsection{Experiment on Dataset D816}

For the experiment D816, the proposed method clearly outperformed other algorithms in all test cases, in terms of both registration error and execution time.
Especially, while other algorithms took substantially longer time than the D815 results, the proposed method maintained almost consistent speed, which highlights the proposed method's fast and accurate registration.
We conjecture that texture-less surfaces and repeated structures of the dataset D816, required algorithms of more complex estimations and much longer time.
Thus, it would be safe to say that when compared to other RANSAC, ICP or BA-based registration algorithms, the proposed algorithm was able to produce more accurate registration results in much shorter time, by incorporating the rotation of the camera that is bound by the two axes of pan-tilt servos into the registration process.

\section{Conclusion}

In this paper, we proposed \emph{Axis Bound Registration}, which is an accurate and fast method for registering RGB-D scans of a pan-tilt camera. 
We incorporated rotation axis calibration and camera-servo control to register point clouds of two frames.
Utilizing the prior knowledge on the rotational motion of the pan-tilt servos, a distance-to-rotation-trajectory constraint was introduced to robustly reject falsely matched pairs without iterative RANSAC process.
The rotational model is used to construct an objective function that minimizes the point-to-point errors for transform estimation. 
The alternating optimization scheme is adopted to divide and linearlize the objective function into two independent problems, which accelerates the solver.
In the experiments, the proposed method was compared with, and outperformed four other ICP or BA-based state-of-the-art registration algorithms, in terms of RMSE and execution time. 

\bibliographystyle{plainnat}
\bibliography{bibsample}

\end{document}